\documentclass[runningheads]{llncs}
\usepackage{graphicx}

\usepackage{pgfplots}
\usepackage{tikz}
\usepackage{mathrsfs}
\usepackage{amsfonts}
\usepackage[ruled]{algorithm2e}
\usepackage{floatrow}
\usepackage{multirow}
\newfloatcommand{capbtabbox}{table}[][\FBwidth]

\SetCommentSty{mycommfont}

\begin{document}
	%
	
	\title{Brain Tumor Segmentation Network Using Attention-based Fusion and Spatial Relationship Constraint}
	\titlerunning{Multi-Modal Brain Tumor Segmentation Network}
	%
	\author{Chenyu Liu \inst{1} \and 
	    Wangbin Ding \inst{1} \and
		Lei Li\inst{2,3,4} \and
		Zhen Zhang \inst{1} \and
		Chenhao Pei \inst{1} \and
		Liqin Huang\inst{1,*} \and
		Xiahai Zhuang\inst{2,*}}
	%
	\authorrunning{Liu et al.}
	%
	\institute{College of Physics and Information Engineering, Fuzhou University, Fuzhou, China 
		\and School of Data Science, Fudan University, Shanghai, China
		\and School of Biomedical Engineering, Shanghai Jiao Tong University, Shanghai, China  
		\and School of Biomedical Engineering and Imaging Sciences, King’s College London, London, UK
	}

	\footnotetext{* L Huang and X Zhuang are co-senior and corresponding authors: hlq@fzu.edu.cn; zxh@fudan.edu.cn. This work was funded by Fujian Science and Technology Project (Grant No. 2019Y9070, 2020J01472),
	National Natural Science Foundation of China (Grant No. 61971142), Shanghai Municipal Science and Technology Major Project (Grant No. 2017SHZDZX01).}
	%
	\maketitle              
	\begin{abstract}

    Delineating the brain tumor from magnetic resonance (MR) images is critical for the treatment of gliomas. However, automatic delineation is challenging due to the complex appearance and ambiguous outlines of tumors. Considering that multi-modal MR images can reflect different tumor biological properties, we develop a novel  multi-modal tumor segmentation network (MMTSN) to robustly segment brain tumors based on multi-modal MR images. The MMTSN is composed of three sub-branches and a main branch. Specifically, the sub-branches are used to capture different tumor features from multi-modal images, while in the main branch, we design a spatial-channel fusion block (SCFB) to effectively aggregate multi-modal features. Additionally, inspired by the fact that the spatial relationship between sub-regions of the tumor is relatively fixed, e.g., the enhancing tumor is always in the tumor core, we propose a spatial loss to constrain the relationship between different sub-regions of tumor. We evaluate our method on the test set of multi-modal brain tumor segmentation challenge 2020 (BraTs2020). The method achieves 0.8764, 0.8243 and 0.773 Dice score for the whole tumor, tumor core and enhancing tumor, respectively.

	\end{abstract}
	\section{Introduction}
    
     Gliomas are malignant tumors that arise from the canceration of glial cells in the brain and spinal cord \cite{Wen2008Malignant}.
    It is a dangerous disease with high morbidity, recurrence and mortality. The treatment of gliomas is mainly based on resection. Therefore, accurate brain tumor segmentation plays an important role in disease diagnosis and therapy planning \cite{bakas2017advancing}. 
    However, automatic tumor segmentation is still challenging, mainly due to the diverse location, appearance and shape of gliomas.
    
    The multi-modal magnetic resonance (MR) images can provide complementary information for the anatomical structure. It  has been largely used for clinical applications, such as brain, heart and intervertebral disc segmentation \cite{zhang2015deep,zhuang2018multivariate,li20183d}. As reported in \cite{menze2015multimodal}, T2 weighted (T2) and fluid attenuation inverted recovery (Flair)  images highlight the peritumoral edema, while T1 weighted (T1) and T1 enhanced contrast (T1c) images visualize the necrotic and non-enhancing tumor core, and T1c futher presents the region of the enhancing tumor. Therefore, the application of the multi-modal MR images for brain tumor segmentation has attracted increasing attention.
    
    Most conventional multi-modal brain tumor segmentation approaches are based on classification algorithms, such as support vector machines \cite{2019Automated} and random forests \cite{2014Appearance}. Recently, based on deep neural network (DNN), Havaei et al. proposed a convolutional segmentation network by using 2D multi-modal images \cite{Havaei2017Brain}, but 2D convolutions can not fully leverage the 3D contextual information.
    Kamnitsas et al. proposed a multi-scale 3D CNN which can perform brain tumor segmentation by processing 3D volumes directly \cite{kamnitsas2017efficient}. Compared to the state-of-the-art 3D network, their model can incorporate both local and larger contextual information for segmentation. Additionally, they utilized a fully connected conditional random fields as the post-processing to refine the segmentation results.
    According to the hierarchical structure of the tumor regions, Wang et al. decomposed the multiple class segmentation task into three cascaded sub-segmentation tasks and each of the sub tasks is resolved by a 3D CNN \cite{wang2017automatic}.  Furthermore, Chen et al. proposed a end-to-end cascaded network for multi-label brain tumor segmentation \cite{2018Focus}. However, such a cascaded method ignored the correlation among the tasks. To tackle this, Zhou et al. \cite{zhou2018one} presented a multi-task segmentation network. They jointly optimized multiple class segmentation tasks in a single model to exploit their underlying correlation.
    In this work, we develop a fully automatic brain tumor segmentation method based on 3D convolution neural network, which can effectively fuse complementary tumor information from multi-modal MR images.
    The main contributions of our method are summarized as follows:
 
    (1) We propose a novel multi-modal tumor segmentation network (MMTSN), and evaluate it on the multi-modal brain tumor segmentation challenge 2020 (BraTs2020) dataset \cite{bakas2018identifying,bakas2017segmentation1,bakas2017segmentation,bakas2017advancing,menze2015multimodal}. 
    
    (2) We propose a fusion block based on spatial and channel attention, which can effectively aggregate multi-modal features for segmentation tasks.
    
    (3) Based on our network, we design a spatial constraint loss. The loss regularizes the spatial relationship of the sub-regions of tumor and improves the segmentation performance.

	\section{Method}
	\subsection{Multi-modal Tumor Segmentation Network}
	\label{sec:reg}
    
\begin{figure}[htb] 
		\centering
		\includegraphics[width=0.94\textwidth]{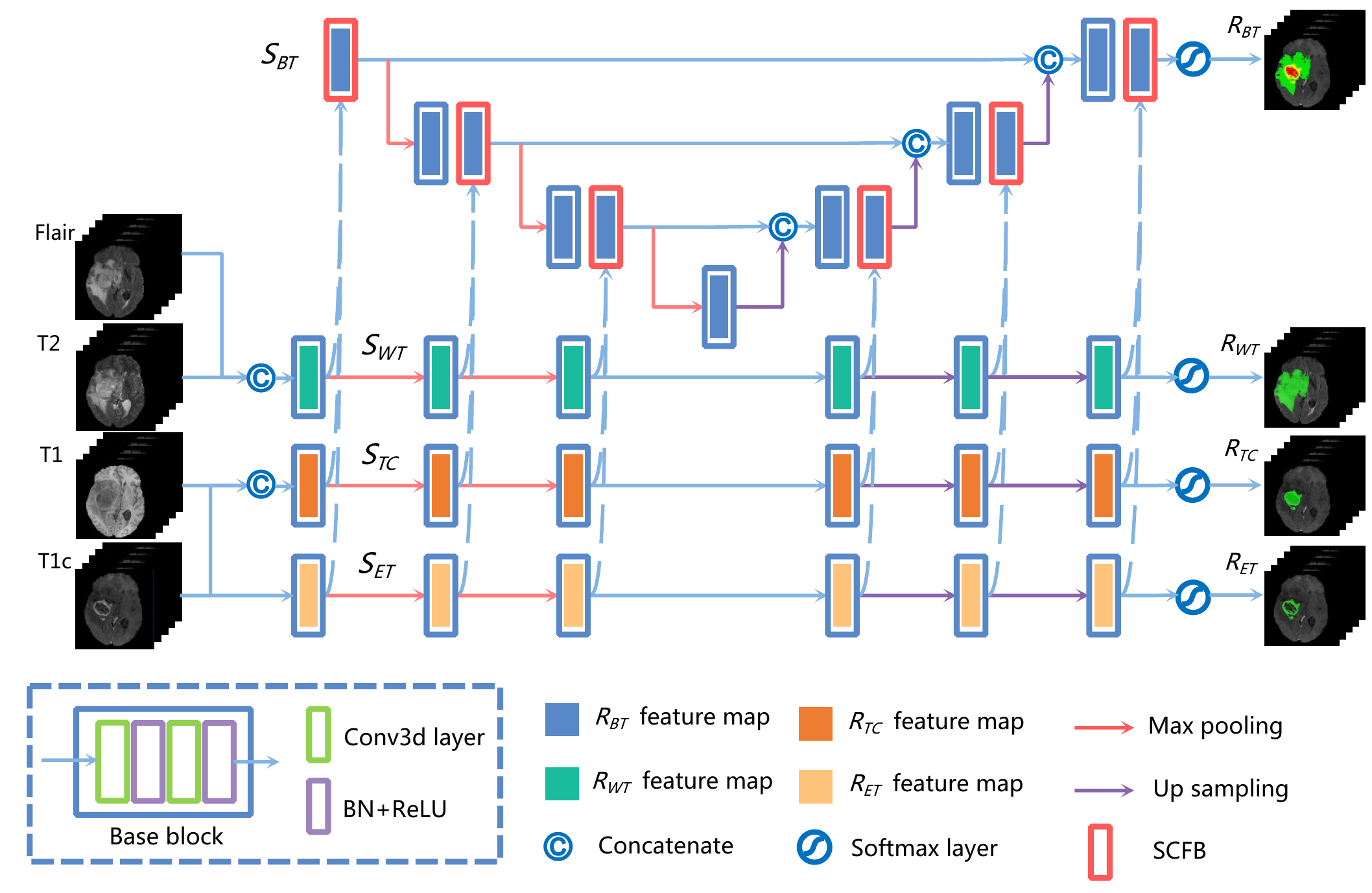}
		\caption{Overview of the MMTSN architecture. The network contains three sub-branches to capture modality-specific information, and a main the branch to effectively fuse multi-modal features for tumor segmentation.}
		\label{fig:1}    
	\end{figure}

    Multi-modal MR images can provide different biological properties of tumor. We propose a MMTSN to fully capture this modality-specific information. Figure \ref{fig:1} shows the architecture of the MMTSN. It is composed of three sub segmentation branches $(S_{WT},S_{TC},S_{ET})$ and a main segmentation branch ($S_{BT}$).
	
    Given a multi-modal MR image $I_{mul}=(I_{T1},I_{T1c},I_{T2},I_{Flair})$, the $S_{WT}$ is used to capture the whole tumor region (WT) by $I_{T2}$ and $I_{Flair}$ images; the $S_{TC}$ aims to acquire tumor core region (TC) by $I_{T1}$ and $I_{T1c}$ images; and the $S_{ET}$ is intent to extract enhanced tumor region (ET) by $I_{T1c}$ image. Therefore, the loss functions of the three branches are defined as
		\begin{equation}
		\mathcal{L}oss_{WT}=1-\mathcal{D}ice(L_{WT},\hat{L}_{WT}),
		\end{equation}
		\begin{equation}
		\mathcal{L}oss_{TC}=1-\mathcal{D}ice(L_{TC},\hat{L}_{TC}),
		\end{equation}
		\begin{equation}
		\mathcal{L}oss_{ET}=1-\mathcal{D}ice(L_{ET},\hat{L}_{ET}),
		\end{equation}
	where $\mathcal{D}ice(A,B)$ calculates the Dice score of $A$ and $B$, ($L_{WT}$, $L_{TC}$, $L_{ET}$) and ($\hat{L}_{WT}$,  $\hat{L}_{TC}$, $\hat{L}_{ET}$ ) are corresponding gold standard and predicted label of regions (WT, TC, ET), respectively.
    
    Having the sub-branches constructed, the multi-modal feature maps in  $(S_{WT},$ $S_{TC}, S_{ET})$  can be extracted and propagated to $S_{BT}$ for segmentation. The backbone of the $S_{BT}$ is in U-Shape  \cite{ronneberger2015u}. To effectively fuse complementary information, we also design a spatial-channel attention based fusion block (see \ref{sec:abfb} for details) for multi-modal feature aggregation. The $S_{BT}$ jointly performs edema, enhancing and non-enhancing$\&$necrotic regions segmentation,
    and the loss function is 
    	\begin{equation}
    	\mathcal{L}oss_{BT}=1-\mathcal{D}ice(L_{BT},\hat{L}_{BT}),
    	\end{equation}
	where $L_{BT}$ and $\hat{L}_{BT}$ are the gold standard and predicted label of all sub-regions of the tumor, respectively. Finally, the overall loss function of the network is 
    	\begin{equation}
    	\mathcal{L}oss_{MMTSN}=\mathcal{L}oss_{BT}+\lambda_{WT}\mathcal{L}oss_{WT}+\lambda_{TC}\mathcal{L}oss_{TC}+
    	\lambda_{ET}\mathcal{L}oss_{ET}+\lambda_{SC}\mathcal{L}oss_{SC},
    	\label{equ:mmtsn}
    	\end{equation}
    where $\lambda_{WT}$, $\lambda_{TC}$, $\lambda_{ET}$ and $\lambda_{SC}$ are hyper-parameters, and the $\mathcal{L}oss_{SC}$ is the spatial constraints loss (see \ref{sec:sclf} for details).

    \subsection{Spatial-Channel Fusion Block (SCFB) }
	\label{sec:abfb}
	We present a spatial-channel attention based fusion block to fuse multi-modal information for segmentation.  According to \cite{chen2020mmfnet}, channel attention can effectively re-calibrate channel-wise feature responses, while spatial attention highlights region of interest. Therefore, combining channel and spatial attention in our fusion block can emphasize feature maps and interest regions for the tumor.
	\begin{figure}[htb] 
		\centering
		\includegraphics[width=0.94\textwidth]{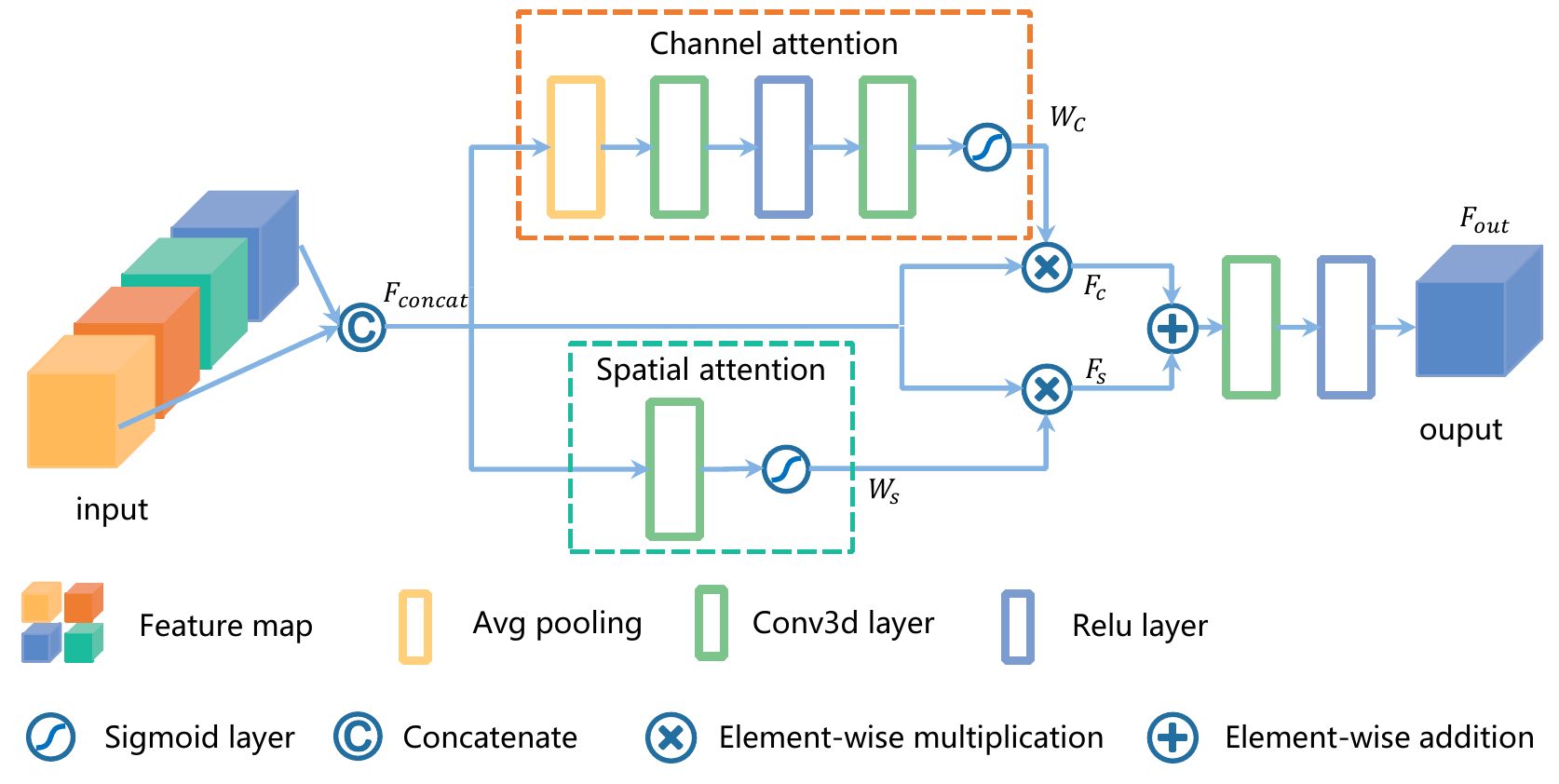}
		\caption{The spatial-channel attention based fusion block.}
		\label{fig:2}    
	\end{figure}
	
	The design of SCFB is shown in Figure \ref{fig:2}. Assume that we have three feature maps (${F_{WT}}$, ${F_{TC}}$, ${F_{ET}}$) from $(S_{WT},S_{TC},S_{ET})$ and one previous output ${F_{BT}}$ from the $S_{BT}$. The SCFB first concatenate  (${F_{WT}}$, ${F_{TC}}$, ${F_{ET}}, {F_{BT}}$) to obtain ${F_{concat}}$. Then, channel attention and spatial attention are applied to both emphasize informative feature maps and highlight interest regions of ${F_{concat}}$. 
	 In the SCFB, the channel attention can be defined as 
	\begin{equation}
		{F_{c}} = {W_{c}} \odot {F_{concat}},
	\end{equation}
    \begin{equation}
		{W_{c}} = \sigma (k^{1 \times 1 \times 1} \alpha (k^{1 \times 1 \times 1} AvgPool({F_{concat}}))),
	\end{equation}
	where ${F_{c}}$ is the output feature maps of the channel attention block, ${W_{c}}$ is the channel-wise attention weight and $\odot$ is the element-wise multiplication, $k^{a \times b \times c}$ is defined as a convolutional layer with a kernel size of ${a \times b \times c}$, $\alpha$ and $\sigma$ is a ReLU layer and sigmoid activation respectively. Meanwhile, the spatial attention can be formulated as 
	\begin{equation}
		{F_{s}} = {W_{s}} \odot {F_{concat}},
	\end{equation}
	\begin{equation}
		{W_{s}} = \sigma (k^{1 \times 1 \times 1} {F_{concat}}),
	\end{equation}
	where ${F_{s}}$ is defined as output feature maps of the spatial attention block and  ${W_{c}}$ is the spatial-wise attention weight. Finally, we combine the output feature maps of channel attention block and spatial attention block by add operation. Therefore, the final output of the SCFB is
	\begin{equation}
		{F_{out}} = \alpha (k^{3 \times 3 \times 3} ({F_{c}} + {F_{s}})).
	\end{equation}

	\subsection{Spatial Relationship Constraint}
	\label{sec:sclf}
\begin{figure}[htb] 
		\centering
		\includegraphics[width=0.7\textwidth]{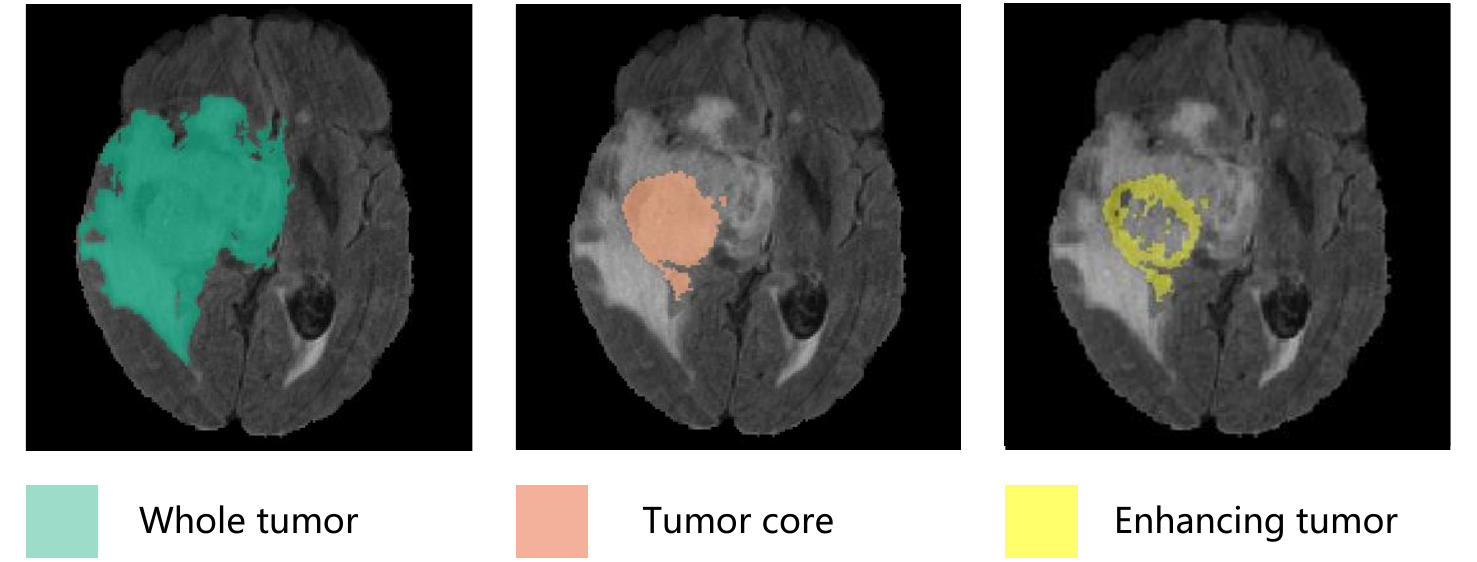}
		\caption{Spatial relationship of different sub-regions in glioma}
		\label{fig:3}    
	\end{figure}

		As shown in Figure \ref{fig:3}, there are spatial relationship between different sub-regions of tumor, i.e, TC is in WT, and the TC contains ET. 
		Thus, we adopt these relationships as spatial constraints (SC) to regularize the segmentation results of MMTSN.
	
	In section \ref{sec:reg}, we have constructed three sub-branches (see Figure \ref{fig:1}) to predict the WT, TC and ET from different MR images separately. The spatial constraint can be formulated based on the prediction result of each branch, 
	\begin{equation}
	\mathcal{L}oss_{SC}^{WT,TC}=1-{\frac{\sum\limits_{x \in\Omega}\hat{L}_{WT}(x)\cdot\hat{L}_{TC}(x)}
		{\sum\limits_{x \in \Omega}\hat{L}_{TC}(x)}},
	\end{equation}
	\begin{equation}
	\mathcal{L}oss_{SC}^{TC,ET}=1-{\frac{\sum\limits_{x \in\Omega}\hat{L}_{TC}(x)\cdot\hat{L}_{ET}(x)}
		{\sum\limits_{x \in \Omega}\hat{L}_{ET}(x)}},
	\end{equation}
	where the $\Omega$ is the common spatial space. Ideally, the $\mathcal{L}oss_{SC}^{WT,TC}$ (or $\mathcal{L}oss_{SC}^{TC,ET}$) is equeal to $0$ when the WT (or TC) completely contains TC (or ET). Finally, the total spatial constraint loss is 
		\begin{equation}
	\mathcal{L}oss_{SC}=\mathcal{L}oss_{SC}^{WT,TC}+
							\mathcal{L}oss_{SC}^{TC,ET}.
	\label{equa:src}
	\end{equation}
	The auxiliary $\mathcal{L}oss_{SC}$ enforces consistent spatial relationship between the sub-branches, so that the feature maps of each sub-branch can retain more accurate spatial information to improve the segmentation performance in the main branch.

	\section{Experiment}
    \subsection{Dataset}
    We used the multi-modal BraTs2020 dataset to evaluate our model. The training set contains images $I_{mul}$ from 369 patients, and the validation set contains images $I_{mul}$ from 125 patients without the gold standard label.
    Each patient was scanned with four MRI sequences: T1, T1c, T2 and Flair, where each modality volume is of size $240 \times 240 \times 155$.
    All the images had already been skull-striped, re-sampled to an isotropic ${1mm^3}$ resolution, and co-registered to the same anatomical template.
    
    \subsection{Implementations}
    Our network was implemented in PyTorch, and trained on NVIDIA GeForce RTX 2080 Ti GPU. In order to reduce memory consumption, the network processed an image patch-wisely. For each $I_{mul}$, we normalized intensity values, and extracted multi-modal patches $P_{mul}=(P_{T1},P_{T1c},P_{T2},P_{Flair})$ with a size of $4 \times 64 \times 64 \times 48$ from it by sliding window technique. Then the patches can be feed into the network for training and testing. Additionally, the gamma correction, random rotation and random axis mirror flip are adopted for data augmentation to prevent overfitting during model training.
    The hyper-parameter in $\lambda_{WT}$, $\lambda_{ET}$, $\lambda_{TC}$ and $\lambda_{SC}$ were set to  $0.5$, $0.6$ , $0.6$ and $0.5$, respectively (see Eq. \ref{equ:mmtsn}). Finally, the network parameters can be updated by minimizing the  $\mathcal{L}oss_{MMTSN}$ with Adam optimizer (learning rate=0.001).
    
    
    
    \subsection{Results}
\begin{figure}[bth] 
		\centering
		\includegraphics[width=0.85\textwidth]{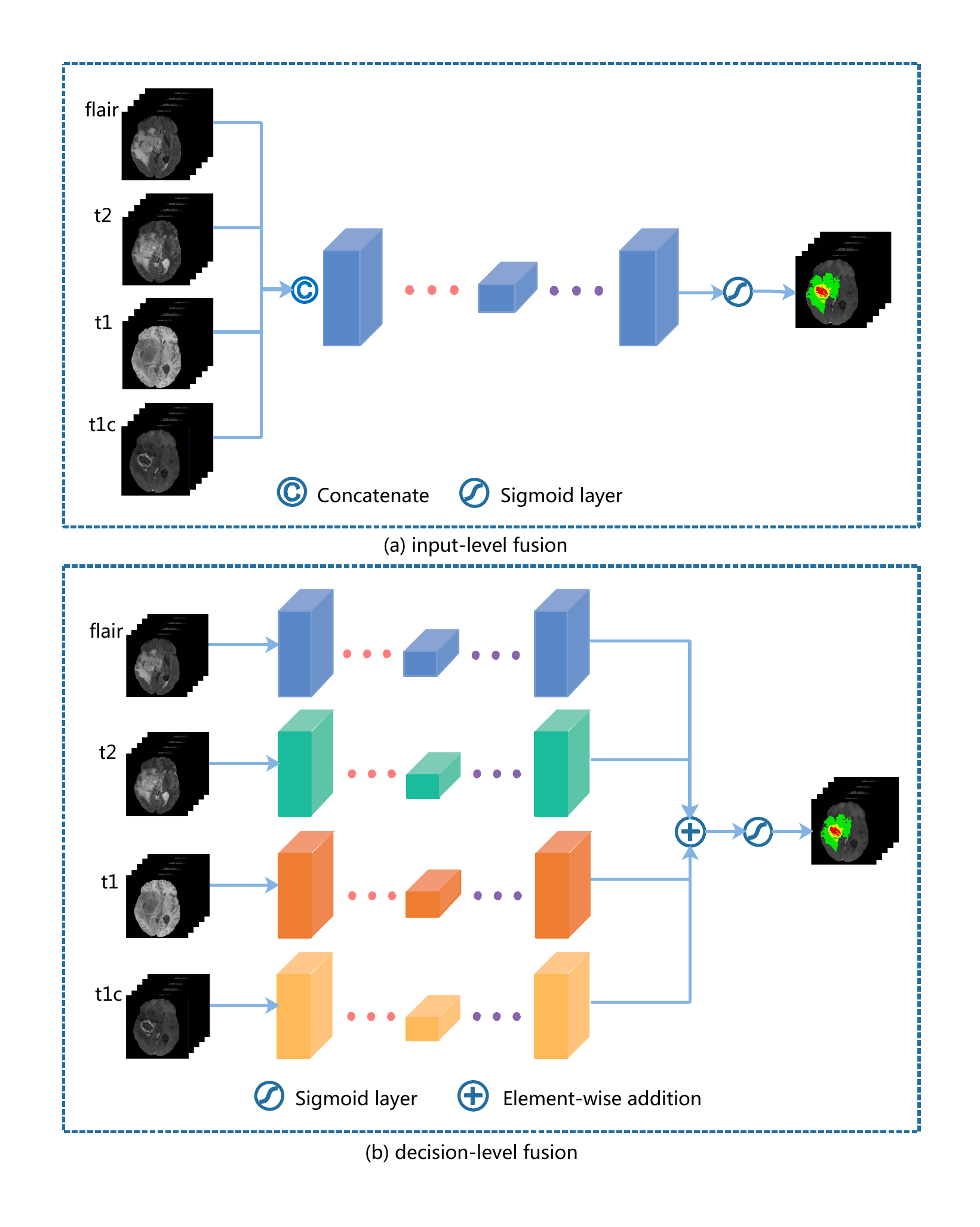}
		\caption{The architecture of two fusion strategies. Input-level fusion directly concatenates multi-modal images as input, while decision-level fusion adds the output of each modality-specific sub-branch to get the final segmentation result. Note that skip connections are not marked, but they are actually involved in both fusion strategies.}
		\label{fig:4}    
	\end{figure}

    To evaluate the performance of our framework, the Dice and 95th percentile of the Hausdorff Distance (HD95) are used as criteria. Table \ref{tab:2} shows the final result of our method on test set.
    Furthermore, To explore the advantage of our network architecture, SCFB module and the SC loss, we conducted to compare our method to five different methods on validation set:
	\begin{itemize}
		\item {3D Unet-pre}: The 3D Unet  which is based on input-level fusion (as shown in Figure \ref{fig:4}(a)) \cite{cciccek20163d}.
		\item {3D Unet-post}: The 3D Unet using decision-level fusion (as shown in Figure \ref{fig:4}(b)) \cite{zhou2019review}.
		\item {MMTSN-WO-SCFB  }: Our MMTSN network but using concatenation rather than SCFB module for feature map fusion.
		\item {MMTSN-WO-$\mathcal{L}oss_{SC}$  }: Our MMTSN network but without SC loss function. 
		\item  MMTSN: Our proposed multi-modal tumor segmentation network. 
	\end{itemize}
    
	 \begin{table*}[ht] 
    	\capbtabbox[0.9\textwidth]{
    		\begin{tabular}{ccccccc}
    			\hline
    			\multirow{2}{*}{} 
    			& \multicolumn{3}{c}{Dice (\%)}  
    			& \multicolumn{3}{c}{HD95 (mm)}  \\
    			\cline{2-7}
    			& ET & TC & WT & ET & TC & WT\\ 
    			\hline
    			Mean
    			& 77.31	 &  82.43   &  87.64	   
    			& 27.17	 &  20.23   &  6.45	  \\
    			Median
    			& 85.00	 &  92.39   &  91.55	   
    			& 1.41	 &  2.45    &  3.16 	\\
    			25 quantile
    			& 75.95	 &  86.08   &  86.49	    
    			& 1.00	 &  1.41    &  2.00	    \\
    			75 quantile
    			& 90.31	 &  95.46   &  94.29	   
    			& 2.83	 &  4.90    &  6.16	    \\
    			
    			\hline
    		\end{tabular}
    	}{
    		\caption{Dice score and HD95 of the proposed method on the test set.}
    		\label{tab:2}
    	}
    \end{table*}

    \begin{table*}[ht] 
    	\capbtabbox[0.9\textwidth]{
    		\begin{tabular}{ccccccc}
    			\hline
    			\multirow{2}{*}{Method} 
    			& \multicolumn{3}{c}{Dice (\%)}  
    			& \multicolumn{3}{c}{HD95 (mm)}  \\
    			\cline{2-7}
    			& ET & TC & WT & ET & TC & WT\\ 
    			\hline
    			3D Unet-pre  
    			& 69.79	 &  79.05   &  87.67	   
    			& 45.64	 &  13.48   &  7.04	  \\
    			3D Unet-post  
    			& 71.98	 &  79.27   &  88.22	  
    			&  36.31 &  16.30   &  6.28	  \\
    			MMTSN-WO-SCFB  
    			& 73.86	 &  79.81  &  \textbf{88.80}	   
    			& 30.67	 &  12.60  &  \textbf{6.14} 	    \\
    			MMTSN-WO-$\mathcal{L}oss_{SC}$  
    			& 75.94	 &  79.67  &  87.12	    
    			& 21.89	 &  14.00  &  7.45	    \\
    			MMTSN  
    			& \textbf{76.37}	&  \textbf{80.12}   &  88.23	   
    			& \textbf{21.39} 	&  \textbf{6.68}    &  6.49	    \\
    			
    			\hline
    		\end{tabular}
    	}{
    		\caption{Dice score and HD95 of the proposed method and other baseline methods on the validation set.}
    		\label{tab:1}
    	}
    \end{table*}
   
	 In Table \ref{tab:1}, compared to 3D Unet-pre and 3D Unet-post, our proposed methods (MMTSN-WO-SCFB, MMTSN-WO-$\mathcal{L}oss_{SC}$ and MMTSN)  performed better both in Dice and HD95. Especially in the more challenging areas (TC and ET), the MMTSN achieved the best accuracy among all compared methods. This demonstrates the effectiveness of our designed architecture (see Figure \ref{fig:1}). 
	
	Also in Table \ref{tab:1}, one can be seen that the  MMSTN with SCFB can achieve better results than MMTSN-WO-SCFB on both Dice score and HD95. It shows the advantage of SCFB for multi-modal feature fusion. Meanwhile, compared to MMTSN-WO-$\mathcal{L}oss_{SC}$, although MMTSN had no obvious improvement in Dice score, it greatly performed better in HD95 criterion. This reveals that SC loss can effectively achieve spatial constraints for segmentation results.  
	  
\begin{figure}[htb] 
		\centering
		\includegraphics[width=0.9\textwidth]{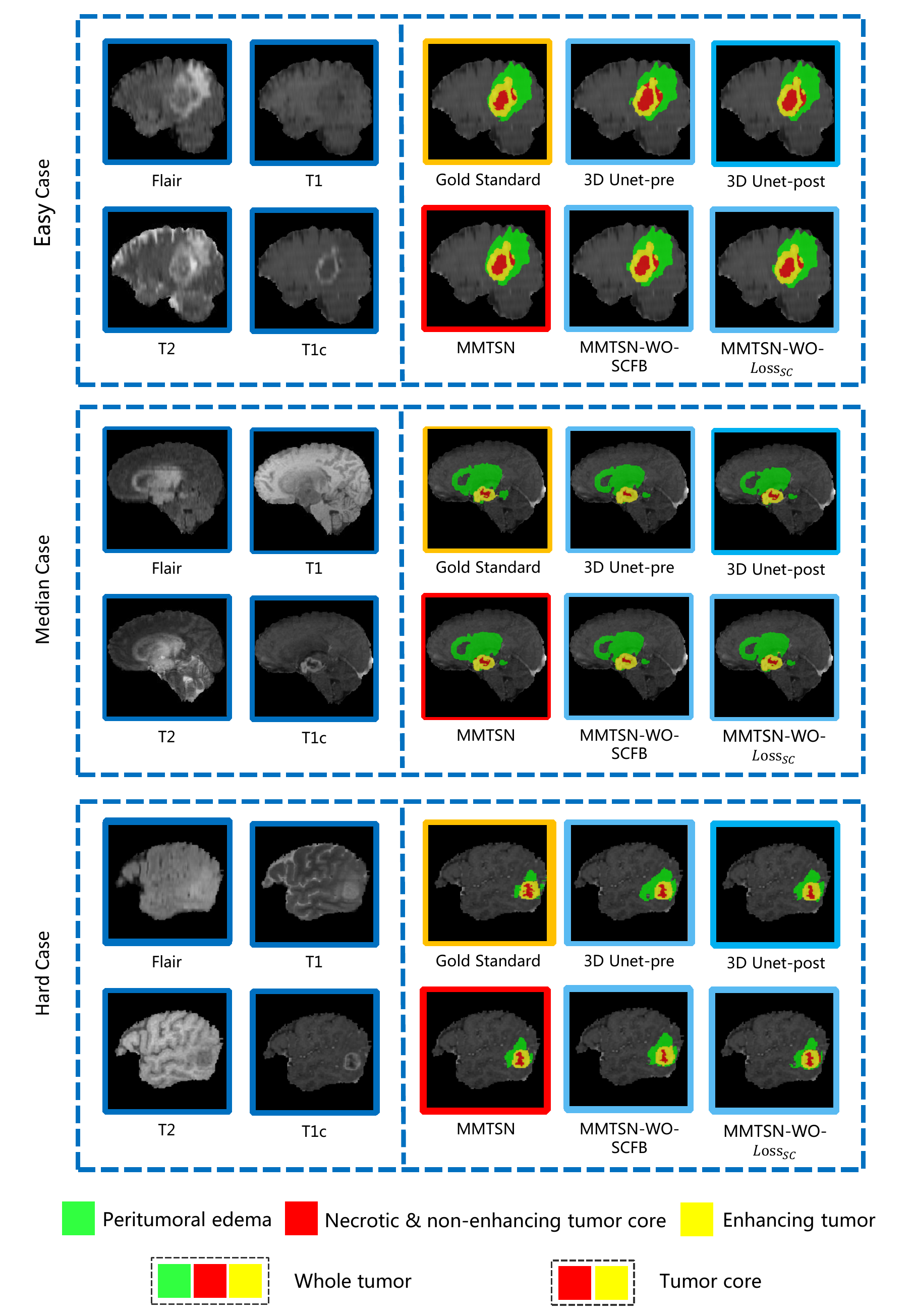}
		\caption{Axial view of three validation cases: the easy, medium and hard case among the validation set, respectively. Our method MMTSN incorporated with SCFB and SC can achieve the best visual result. 
		}
		\label{fig:result}    
	\end{figure}

	Additionally, Figure \ref{fig:result} shows the visual results of three different cases. For the edema region segmentation (green), even though all of the methods obtained comparable results in the easy and median case, the MMTSN still showed potential advantages in the hard case. For enhancing tumor segmentation (yellow), one can see that the MMTSN and MMTSN-WO-$\mathcal{L}oss_{SC}$ performed better than other methods, which is consistent with the quantitative result in Table \ref{tab:1}. 
	For the challenging necrotic and non-enhancing segmentation (red), the figure suggests that the MMTSN can obtain relatively better visual results among all the cases.
	%
	
	
	\section{Conclusion}
    In this work, we proposed a 3D MMTSN for brain tumor segmentation. We constructed three sub-branches and a main branch to capture modality-specific and multi-modal features. In order to fuse useful information of different MR images, we introduced a spatial-channel attention based fusion block. Furthermore, a spatial loss was designed to constrain the relationship between different sub-regions of glioma. We evaluated our method on the multi-modal BraTs2020 dataset to demonstrate the effectiveness of the MMTSN framework. Future work aims to apply our method to other medical image segmentation scenarios.

\bibliographystyle{splncs04}

\bibliography{ref}

\end{document}